\documentclass[twocolumn]{revtex4-1}

\usepackage[utf8]{inputenc}
\usepackage[english]{babel}
\usepackage{csquotes}
\usepackage{graphicx}
\usepackage[separate-uncertainty=true]{siunitx}
\usepackage[colorlinks,allcolors=black,citecolor=blue,urlcolor=blue]{hyperref}
\usepackage{textcomp}
\usepackage{gensymb}

\graphicspath{{figures/}}

\begin{document}

\title{Quantum dynamics and spectroscopy of ab initio liquid water:\\the interplay of nuclear and electronic quantum effects}

\author{Ondrej Marsalek}
\author{Thomas E. Markland}
\email{tmarkland@stanford.edu}
\affiliation{Department of Chemistry, Stanford University, Stanford, California 94305, USA}

\begin{abstract}
Understanding the reactivity and spectroscopy of aqueous solutions at the atomistic level is crucial for the elucidation and design of chemical processes.
However, the simulation of these systems requires addressing the formidable challenges of treating the quantum nature of both the electrons and nuclei.
Exploiting our recently developed methods that provide acceleration by up to two orders of magnitude, we combine path integral simulations with on-the-fly evaluation of the electronic structure at the hybrid density functional theory level to capture the interplay between nuclear quantum effects and the electronic surface.
Here we show that this combination provides accurate structure and dynamics, including the full infra-red and Raman spectra of liquid water.
This allows us to demonstrate and explain the failings of lower-level density functionals for dynamics and vibrational spectroscopy when the nuclei are treated quantum mechanically.
These insights thus provide a foundation for the reliable investigation of spectroscopy and reactivity in aqueous environments.
\end{abstract}

\maketitle

The ability to perform simulations that allow for the making and breaking of bonds in aqueous environments is essential for the description of a wide range of chemical processes.
Ab initio molecular dynamics (AIMD) simulations achieve this by performing dynamics where the interactions are evaluated on the fly from electronic structure calculations. These simulations provide access to a wide variety of dynamical properties, ranging from diffusion coefficients and reorientation times to reaction rates and spectroscopy. However, obtaining the relevant time correlation functions to converge these properties in solution typically requires simulations of hundreds of picoseconds. Hence, with the conventionally required time steps of $\sim$0.5~fs, hundreds of thousands of electronic structure calculations are needed. This cost further increases dramatically if one also wants to treat nuclear quantum effects (NQEs), which have been shown to decrease the water band gap by up to 0.7 eV~\cite{DelBen2015/10.1063/1.4927325,Chen2016/10.1103/PhysRevLett.117.186401}, increase acid dissociation constants in enzyme active sites by almost four orders of magnitude~\cite{Wang2014/10.1073/pnas.1417923111}, and are essential for simulating atmospheric isotope separation processes~\cite{Markland2012/10.1073/pnas.1203365109}.
NQEs can be included exactly for static properties and approximately for dynamics using path integral methods that map the problem onto a series of replicas of the classical system connected by harmonic springs. The centroid molecular dynamics (CMD)~\cite{Cao1994/10.1063/1.468400,Jang1999/10.1063/1.479515} and ring polymer molecular dynamics (RPMD)~\cite{Craig2004/10.1063/1.1777575,Habershon2013/10.1146/annurev-physchem-040412-110122} approaches employ this framework to approximate dynamical properties.
However, 30--100 replicas are typically needed for aqueous systems at room temperature, increasing the total number of electronic structure calculations well into the millions.

This cost has severely limited the ability to obtain dynamical properties from classical AIMD simulations and completely prevented their convergence in path integral AIMD simulations of condensed phase aqueous systems.
The problem has been compounded by the limitations of the generalized gradient approximation (GGA) density functionals, the most common choice for condensed phase AIMD simulations, as well as by technical issues~\cite{Gillan2016/10.1063/1.4944633}.
GGA functionals have been observed to produce aqueous solutions that are overstructured and exhibit glassy dynamics.
However, recent work has shown that these issues are significantly alleviated when larger basis sets~\cite{Lee2007/10.1063/1.2718521,DelBen2015/10.1063/1.4927325} and dispersion corrections~\cite{Bankura2014/10.1021/jp506120t} are used.
Additionally, increases in computational power and methodological developments have recently made higher rungs of the density functional theory (DFT) ladder, such as hybrid, meta-GGA, and non-local correlation density functionals, accessible for classical AIMD simulations of liquid water, albeit with considerable computational effort~\cite{Guidon2008/10.1063/1.2931945,Zhang2011/10.1021/ct2000952,Distasio2014/10.1063/1.4893377,Miceli2015/10.1063/1.4905333}.
The accuracy of these functionals in benchmark calculations of hydrogen bonded systems suggests they could provide an improved description of the condensed phases of water~\cite{Santra2009/10.1063/1.3236840,Gillan2016/10.1063/1.4944633}.

Recently, we have shown that one can accelerate path integral AIMD calculations by almost two orders of magnitude~\cite{Marsalek2016/10.1063/1.4941093,Kapil2015/10.1063/1.4941091} by combining ring polymer contraction~\cite{Markland2008/10.1063/1.2953308,Markland2008/10.1016/j.cplett.2008.09.019,Fanourgakis2009/10.1063/1.3216520} (RPC) with a multiple time stepping (MTS) propagation scheme~\cite{Tuckerman1992/10.1063/1.463137}.
These methods use a reference system, which provides a cheap approximation to interactions over short distances and short time scales, to dramatically reduce the number of full electronic structure calculations required to evaluate interactions along the quantum path and to evolve it in time, respectively.
While other choices of reference system are possible~\cite{Anglada2003/10.1103/PhysRevE.68.055701,Guidon2008/10.1063/1.2931945,Luehr2014/10.1063/1.4866176,Geng2015/10.1016/j.jcp.2014.12.007}, in this work we use self-consistent charge density functional tight binding (SCC-DFTB), which allows us to evaluate the full electronic structure on a single replica instead of the whole 32-replica path and to use that evaluation to propagate for a much larger time step of 2~fs. This reduces the total number of full electronic structure calculations needed for an ab initio path integral simulation of hundreds of picoseconds from millions to tens of thousands.

Here we apply these methods to obtain almost nanosecond time scale simulations and thus, for the first time, converged dynamical properties of liquid water with quantum nuclei and the electronic potential energy surface (PES) calculated on the fly at the hybrid density functional theory level.
In the revPBE0-D3 path integral simulations the practical speed-up achieved by using RPC and MTS was over 90 fold, i.e. the same computational resources would have provided less than 4~ps of path integral dynamics using a standard implementation.
In fact, the speed of our path integral simulations was even 2.5 fold faster than would be obtained from traditional (i.e. without MTS) classical AIMD.
These ab initio path integral simulations allow us to calculate the infra-red (IR) and Raman vibrational spectra, reorientation times, and diffusion coefficients, which are all in excellent agreement with experiment.
By doing this, we show that the competition between NQEs in water depends sensitively on the density functional's ability to accurately describe the balance between covalent and hydrogen bonding.

The revPBE GGA exchange-correlation functional~\cite{Zhang1998/10.1103/PhysRevLett.80.890}, which arises from a single parameter change in the exchange enhancement factor in the PBE functional~\cite{Perdew1996/10.1103/PhysRevLett.77.3865}, has recently been suggested to give accurate properties of liquid water when treating the nuclei classically under a wide range of conditions when combined with dispersion corrections~\cite{Remsing2014/10.1021/jz501067w,Bankura2014/10.1021/jp506120t,Skinner2016/10.1063/1.4944935}.
It has also been shown to perform well in benchmark calculations of neutral, protonated and deprotonated water clusters up to large sizes~\cite{Goerigk2011/10.1039/c0cp02984j,Gillan2014/10.1063/1.4903240} and ice polymorphs~\cite{Brandenburg2015/10.1063/1.4916070} when combined with a dispersion correction~\cite{Grimme2010/10.1063/1.3382344}.
These benchmarks also suggest that its hybrid counterpart revPBE0-D3~\cite{Goerigk2011/10.1039/c0cp02984j}, which incorporates 25~\% exact exchange, reduces the error by a further factor of 2 in the mean absolute deviation~\cite{Goerigk2011/10.1039/c0cp02984j}, although to our knowledge this functional has not been previously studied as to its condensed phase water properties.

Even classical ab initio molecular dynamics simulations have typically only been performed on time scales of $\sim$20~ps, which yield statistical error bars of more than 20~\% on dynamical quantities such as the diffusion coefficient (see SI Section 3).
Due to the efficiency of RPC combined with MTS, it is now also possible to include NQEs in the dynamics of water with hybrid functionals using modest computational resources.
Here we perform path integral AIMD simulations with the revPBE-D3 GGA functional for over 700~ps and the revPBE0-D3 hybrid functional for over 300~ps, reducing these errors to below 5~\%. By contrasting these highly converged path integral simulations with over 1 ns of classical AIMD simulations, we can thus unambiguously characterize the nuclear quantum effects for these functionals and how they manifest in the structure and vibrational spectroscopy.

\begin{figure}[ht]
  \centering
  \includegraphics{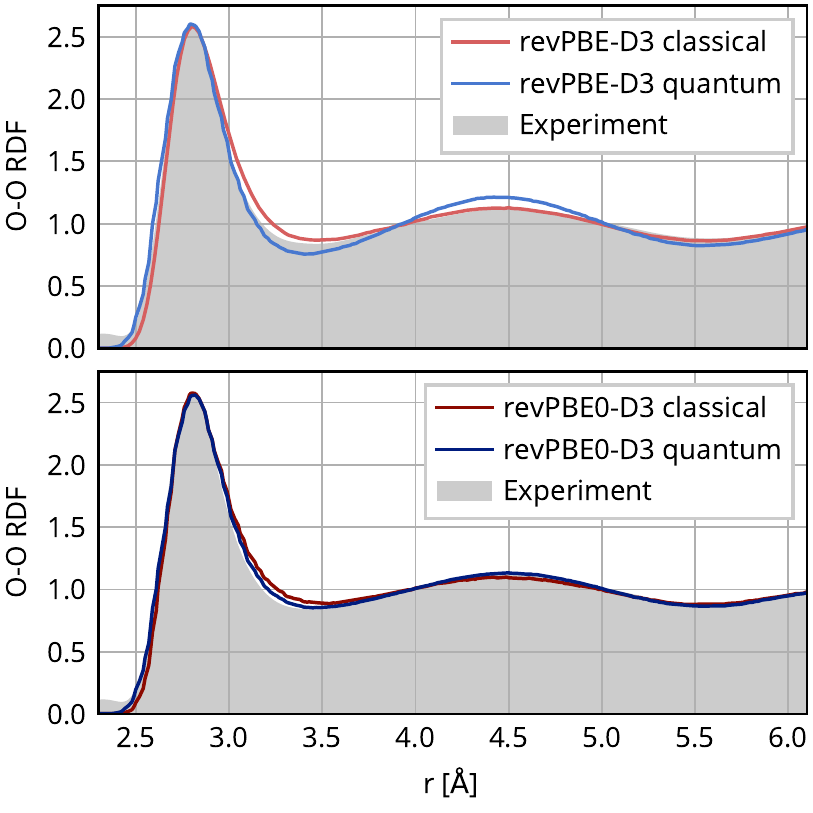}
  \caption{\label{fig:RDF-OO}
    Oxygen-oxygen radial distribution functions.
    Data from classical and path integral simulations at 300~K is shown in the top panel for the revPBE-D3 density functional and in the bottom panel for the revPBE0-D3 functional. The experimental result~\cite{Skinner2013/10.1063/1.4790861} is shown as gray shading for reference.
  }
\end{figure}

\begin{figure}[ht]
  \centering
  \includegraphics{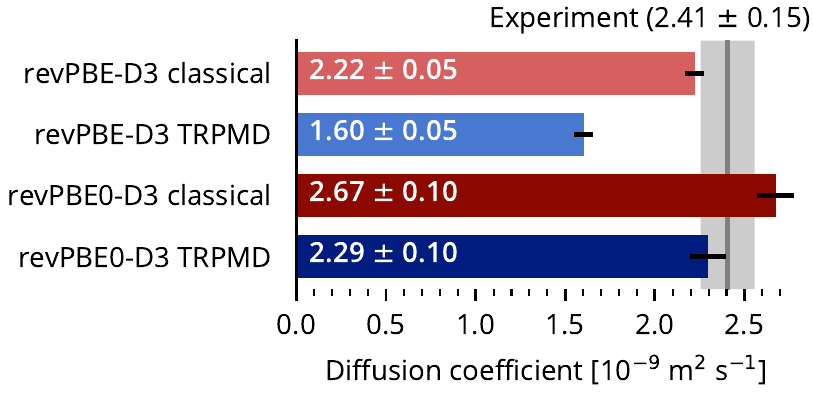}
  \caption{\label{fig:diffusion}
    Diffusion coefficients.
    Water molecule self diffusion coefficients from classical and TRPMD path integral simulations at T=300~K are shown as horizontal bars. Error bars at the 3$\sigma$ confidence level determined using bootstrapping are shown as black lines. The gray line and shading indicate the experimental result and its error bar~\cite{Holz2000/10.1039/b005319h}.
  }
\end{figure}

Figure~\ref{fig:RDF-OO} shows that, consistent with previous simulations~\cite{Skinner2016/10.1063/1.4944935}, classical nuclei revPBE-D3 simulations at \SI{300}{K} yield an oxygen-oxygen radial distribution function (RDF) that is in excellent agreement with the latest X-ray data~\cite{Skinner2013/10.1063/1.4790861}.
In addition, Figure~\ref{fig:diffusion} demonstrates that our 800~ps of classical revPBE-D3 simulation gives a system size corrected diffusion coefficient of \SI{2.22+-0.05}{10^{-9}.m^2.s^{-1}}, within statistical error bars of the experimental value~\cite{Holz2000/10.1039/b005319h} of \SI{2.41+-0.15}{10^{-9}.m^2.s^{-1}}.
Including NQEs leads to a mild structuring of the RDF, consistent with increased local tetrahedrality, as previously observed in other GGA functionals~\cite{Ceriotti2016/10.1021/acs.chemrev.5b00674,Marsalek2016/10.1063/1.4941093,Gasparotto2016/10.1021/acs.jctc.5b01138}.
Owing to the speed increases afforded by RPC, we performed 600~ps of thermostatted RPMD (TRPMD) simulations, which approximately include the quantum dynamics of the nuclei, to obtain a size-corrected value for the quantum diffusion coefficient of \SI{1.6+-0.05}{10^{-9}.m^2.s^{-1}}.
This decrease in the diffusion coefficient of $\sim$30~\% upon including NQEs is consistent with the increased structure observed in the O-O RDF.
A similar slowdown is observed in the orientational correlation times in Table~S1 which again shifts revPBE-D3 properties with quantum nuclei away from experiment.

\begin{figure}[ht]
  \centering
  \includegraphics{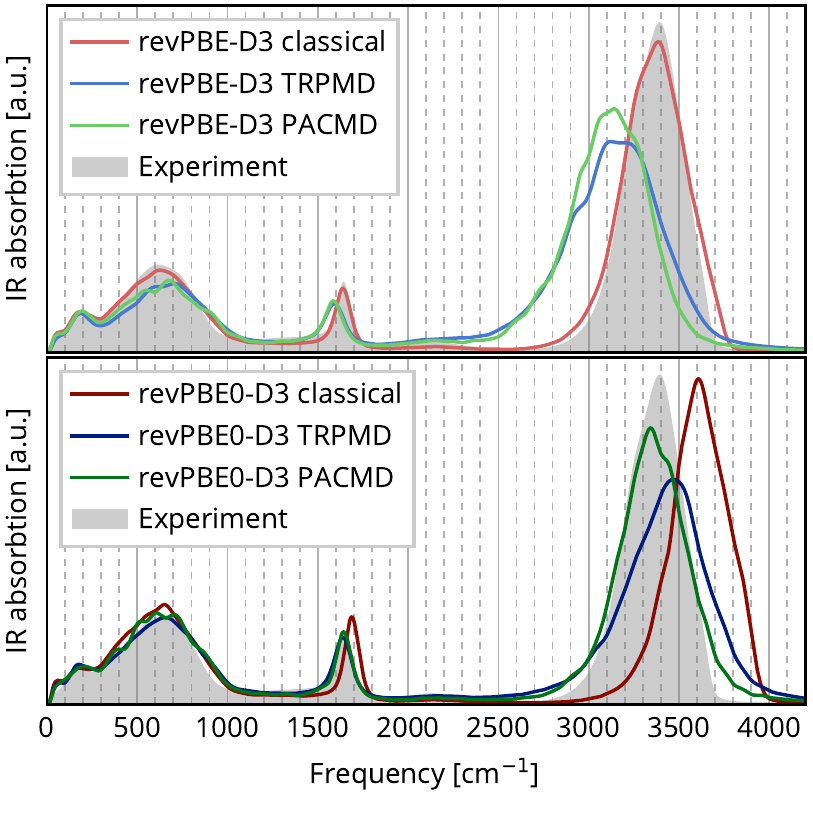}
  \caption{\label{fig:IR}
    Infra-red absorption spectra.
    IR spectra calculated using either classical dynamics or approximate quantum dynamics (TRPMD and PACMD) at T=300~K are shown in the top panel for the revPBE-D3 density functional and in the bottom panel for the revPBE0-D3 functional. The experimental result~\cite{Bertie1996/10.1366/0003702963905385} is shown as gray shading for reference.
  }
\end{figure}

\begin{figure}[ht]
  \centering
  \includegraphics{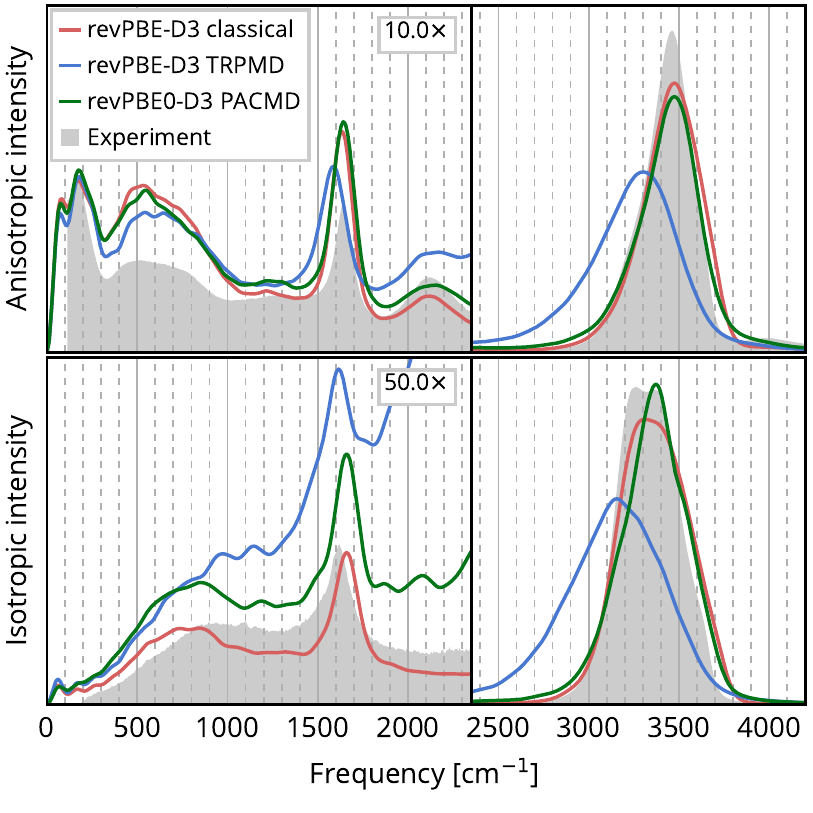}
  \caption{\label{fig:Raman}
    Raman scattering spectra.
    Anisotropic (top panel) and isotropic (bottom panel) Raman spectra obtained with the revPBE-D3 and revPBE0-D3 density functionals and either classical dynamics or approximate quantum dynamics (TRPMD and PACMD) at T=300~K are compared to experimental results~\cite{Raman} shown as gray shading. All the spectra are shown multiplied by the Bose-Einstein correction $1 - \exp(-\beta \hbar \omega) $.
  }
\end{figure}

This failure is most evident in the IR and anisotropic and isotropic Raman vibrational spectra (Figures~\ref{fig:IR} and \ref{fig:Raman}).
While the IR spectrum is given by temporal correlations of the total dipole moment of the system, the Raman spectra are determined by those of the polarizability tensor (see SI Section 7 for details).
These spectroscopies are thus sensitive to different motions in the system, and hence assessing all three provides a broader window on the successes and failures of a computational approach.
While the revPBE-D3 functional with classical nuclei is, remarkably, in essentially perfect agreement with experiment, adding NQEs reveals the classically concealed fundamental shortcomings of the GGA functional.
We computed the vibrational spectra using both TRPMD~\cite{Rossi2014/10.1063/1.4883861} and partially adiabatic CMD (PACMD)~\cite{Hone2006/10.1063/1.2186636,Habershon2008/10.1063/1.2968555}.
Both of these approaches have similar origins and give very similar predictions for diffusion constants and orientational correlation times~\cite{Rossi2014/10.1063/1.4883861} but have established deficiencies when used to compute IR spectra~\cite{Habershon2008/10.1063/1.2968555,Ivanov2010/10.1063/1.3290958,Paesani2010/10.1063/1.3291212}.
However, while TRPMD and PACMD can differ in O-H stretch peak widths for water at 300~K, they have been shown to give consistent quantum effects on IR peak positions~\cite{Rossi2014/10.1063/1.4901214}.
This is shown in the top panel of Figure~\ref{fig:IR}, where for revPBE-D3 both the TRPMD and PACMD spectrum exhibit a large red tail in the O-H stretch region (\SIrange{2500}{4000}{cm^{-1}}) as well as a shift of the H-O-H bend (\SIrange{1500}{1700}{cm^{-1}}) and a distortion of the far IR region (\SIrange{300}{1000}{cm^{-1}}).
This suggests that these changes in the spectra are not artifacts of the quantum dynamics approach.
In addition, these deviations are not removed by using a hybrid functional to calculate the dipole moment using the structures generated from the revPBE-D3 GGA simulations (see Figure S9).
The Raman spectra in Figure~\ref{fig:Raman} show similar features, confirming the deficiencies of the GGA functional.
These results therefore indicate that the remarkable ability of revPBE-D3 to reproduce the structure and dynamics of water in classical simulations arises partially from a fortuitous cancellation between errors in the functional and the neglect of NQEs.
Once NQEs are included, this cancellation breaks down and the deficiencies of the electronic PES become apparent.

As seen in Figure~\ref{fig:RDF-OO} and Figures~S1 and S2, including exact exchange has virtually no effect on the classical structure, consistent with previous observations for the PBE functional~\cite{Guidon2008/10.1063/1.2931945,Guidon2010/10.1021/ct1002225}. However, it does modulate the size of NQEs, decreasing their structuring influence such that the quantum effect on the O-O structure is close to zero for revPBE0-D3. More pronounced changes are seen in the dynamics, where the hybrid diffuses and reorients faster than its GGA counterpart in classical simulations. Once again, the size of NQEs is reduced, with the TRPMD and PACMD simulations giving excellent agreement with the experimental data. In contrast to the GGA case, the simulations including NQEs now exhibit better agreement with experiment than the classical simulations, as should be expected since the nuclei in water are indeed quantum particles. 

The change upon including exact exchange is perhaps most pronounced in the case of the vibrational spectra, where in the classical case it leads to a blue shift of the O-H stretch and, to a lesser extent, the bend region, as shown in Figure~\ref{fig:IR}.
This leads to worse agreement with experiment, which is rectified upon quantizing the nuclei, with both PACMD and TRPMD showing a red shift of the revPBE0-D3 peaks (bottom panels of Figures~\ref{fig:IR} and \ref{fig:Raman}).
The red shift in the IR spectrum upon including NQEs of \SI{\sim 200}{cm^{-1}} in the position of the O-H peak maximum and \SI{\sim 50}{cm^{-1}} in the bend maximum is broadly consistent with those observed previously for water models of varying sophistication~\cite{Habershon2008/10.1063/1.2968555,Liu2009/10.1063/1.3254372,Habershon2009/10.1063/1.3276109,Rossi2014/10.1063/1.4901214}.
Additionally, the distortion of the far IR region is not present with the hybrid functional.
Similar shifts of the O-H peak upon nuclear quantization occur in the Raman spectra in Figure~\ref{fig:Raman}.

The resulting IR and Raman spectra for revPBE0-D3 including NQEs are in good overall agreement with experiment, with PACMD showing a slightly better match in the O-H stretch region of the IR spectrum and TRPMD showing an artificial broadening of the peak, which appears more pronounced than for empirical potentials~\cite{Rossi2014/10.1063/1.4901214}.
Figure~\ref{fig:Raman} shows that this excellent agreement of revPBE0-D3 PACMD continues to the Raman spectrum, where it also captures the bending and libration mode combination band at \SI{2125}{cm^{-1}} in the anisotropic spectrum which is present but less distinct in the IR spectrum.

Is there a structural origin of these vibrational peak shifts? To investigate this, we consider the proton sharing coordinate $\delta = d_{\mathrm{OH}} - d_{\mathrm{O'H}}$, where $d_{\mathrm{OH}}$ and $d_{\mathrm{O'H}}$ are the distances of the proton from the hydrogen bond donor and acceptor oxygen atoms, respectively. Due to the large amount of zero-point energy in the O-H stretch added upon quantization (equivalent to raising the temperature of that coordinate by \SI{\sim 2000}{K}), the distribution along this coordinate is noticeably changed. In particular, there is a large increase in proton sharing in the hydrogen bond, which in extreme cases has been previously suggested to even lead to ``proton excursion'' events~\cite{Ceriotti2013/10.1073/pnas.1308560110}, where the proton is transiently closer to the acceptor than the donor oxygen ($\delta>0$). In our simulations these proton excursions are substantially less common than in previous studies~\cite{Ceriotti2013/10.1073/pnas.1308560110,Wang2014/10.1063/1.4894287}, with none observed with classical nuclei and only 0.005~\% and 0.0009~\% of protons exhibiting them in the GGA and hybrid path integral simulations, respectively. The reasons for these differences are two-fold. Firstly, the functionals employed in those studies all give rise to excessive hydrogen bonding, which is manifested in their overstructured RDFs and sluggish dynamics. Secondly, those studies were performed with the PIGLET approach~\cite{Ceriotti2012/10.1103/PhysRevLett.109.100604} which, by maintaining a non-equilibrium steady state distribution using colored noise, can overestimate extreme fluctuations in the tails of position distributions~\cite{Kapil2016/1606.00920}.

\begin{figure}[ht]
  \centering
  \includegraphics{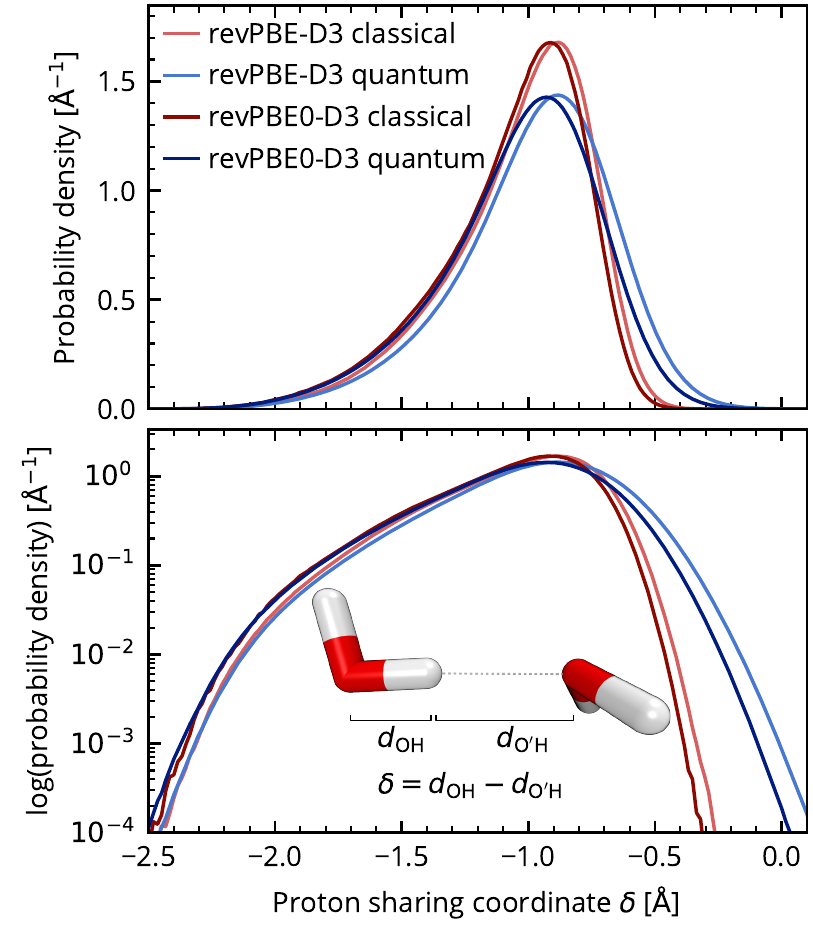}
  \caption{\label{fig:delta}
    Proton sharing in hydrogen bonds.
    Probability distributions of the proton sharing coordinate $\delta$ from classical and path integral simulations at T=300~K are shown in linear (top panel) and logarithmic scale (bottom panel). The definition of the coordinate is shown as an inset. The value is defined for each hydrogen atom in the system using the two nearest oxygen atoms.
  }
\end{figure}

\begin{figure}[ht]
  \centering
  \includegraphics{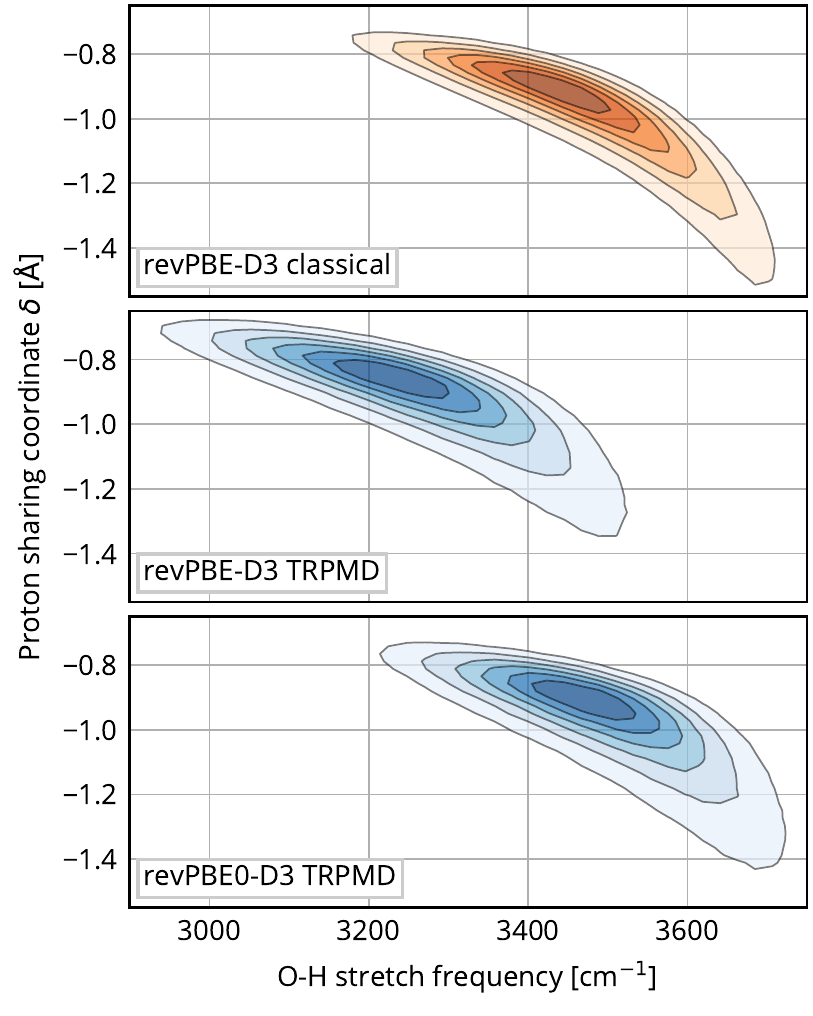}
  \caption{\label{fig:omega-delta}
    Relationship of proton sharing and vibrational frequencies.
    Correlation of the instantaneous frequency of the O-H stretch and the corresponding proton sharing coordinate $\delta$ for revPBE-D3 classical MD (top panel), revPBE-D3 TRPMD (middle panel), and revPBE0-D3 TRPMD (bottom panel) at T=300~K, is shown as a two-dimensional probability distribution. We use a Hann window of total width 300~fs to obtain the instantaneous frequency and convolve the value of the $\delta$ coordinate, as detailed in the SI.
  }
\end{figure}

As shown in Fig.~\ref{fig:delta}, even though NQEs do not lead to a large number of excursions, they still increase proton sharing, with the GGA showing a larger effect than the hybrid.
Since the low frequency side of the O-H vibrational band is typically associated with strong hydrogen bonds, one might expect these highly shared protons to lead to low frequency O-H stretches.
Figure~\ref{fig:omega-delta} shows the instantaneous O-H stretch frequency, computed from the vibrational density of states (VDOS) of the proton, plotted against its proton sharing coordinate $\delta$ (see SI Section 5 for details).
In all cases the frequency of that proton does indeed decrease as the proton sharing increases ($\delta \to 0$).
The exact exchange in the revPBE0-D3 hybrid functional tightens the O-H covalent bond, leading to less proton sharing and hence shifting the O-H frequency distribution back to the correct range. Fortuitously, the revPBE-D3 GGA functional gives a very similar frequency-$\delta$ correlation when used with classical nuclei, albeit for the wrong reasons --- the erroneously loose O-H covalent bonds compensate for the lack of NQEs.

In conclusion, utilizing our latest developments it is now possible to perform path integral simulations long enough to isolate the effects of nuclear quantization on the dynamics and structure of ab initio liquid water.
Our results demonstrate that NQEs can and indeed should be routinely included in ab initio simulations of 100's of atoms on time scales approaching nanoseconds.
The simulations have allowed us to uncover the interplay between the features of the DFT electronic PES and the motion of the quantum nuclei on that PES. In particular, the overall strength of hydrogen bonding and the resulting dynamics sensitively depend on the balance between covalent bonding and proton sharing. This balance is disrupted in the GGA functional when NQEs are included, which manifests as a slight structuring and, importantly, a substantial slowing down of dynamics and distortion of vibrational spectra. This work thus explicitly exposes how the shortcomings of GGA functionals manifest in the properties of liquid water when NQEs are included. This has been hinted at in benchmarking studies where GGA functionals were shown to give larger errors in relative potential energies than hybrids when path integral, rather than classical configurations were used~\cite{Morales2014/10.1021/ct500129p,Ceriotti2016/10.1021/acs.chemrev.5b00674}. On the other hand, the revPBE0-D3 hybrid functional PES correctly balances the NQEs on hydrogen bonding, yielding the excellent agreement we see between our simulations and experiment.

These results add to the rapidly emerging picture of competing quantum effects on the hydrogen bond, which were first suggested based on simple empirical models with anharmonic O-H stretch terms~\cite{Habershon2009/10.1063/1.3167790}, but have since been extended to explain a wide range of effects~\cite{Markland2012/10.1073/pnas.1203365109,Li2011/10.1073/pnas.1016653108,McKenzie2014/10.1063/1.4873352,Wang2014/10.1063/1.4894287,Fang2016/10.1021/acs.jpclett.6b00777,Ceriotti2016/10.1021/acs.chemrev.5b00674,Cendagorta2016/10.1039/C6CP05968F}.
Our use of advanced electronic surfaces generated on the fly now makes it abundantly clear that many static and dynamical properties of bulk liquid water at ambient conditions are determined by an almost perfect cancellation of quantum effects.
For example, the NQE on the diffusion coefficient of revPBE0-D3 corresponds to the value obtained from just a 7~K decrease in temperature (see SI Section 3).
This is in contrast with earlier studies suggesting much larger NQEs on diffusion, with increases by up to a factor of 2~\cite{Poulsen2005/10.1073/pnas.0408647102,Miller2005/10.1063/1.2074967,delaPena2004/10.1063/1.1783871,delaPena2006/10.1063/1.2238861,Paesani2006/10.1063/1.2386157,Liu2009/10.1063/1.3254372,Habershon2009/10.1063/1.3167790}.
Indeed, recently revised diffusion coefficients obtained for the MB-pol PES, which was fit to high-level ab initio calculations, now also similarly show close to zero NQEs~\cite{Reddy2016/1609.02884}.
However, it is clear that one cannot expect this level of compensation in general --- at other thermodynamic state points, at interfaces or in the presence of other hydrogen bonding species.

In summary, by using path integral simulations to include NQEs, we have shown that it is possible to obtain accurate structure, dynamics and spectroscopy of liquid water from hybrid density functionals. In particular, both NQEs and exact exchange have a pronounced yet opposite effect on the dynamics and vibrational spectra of water, while having only a mild effect on most structural properties. Only a combination of both of these effects together in the revPBE0-D3 path integral simulations results in a balanced description and properties in excellent agreement with experiment. This combination of an accurate electronic PES and treatment of NQEs, which is now computationally tractable by exploiting multiple time scale techniques, thus offers the opportunity to investigate dynamics and spectroscopy as well as reactivity in aqueous systems in a wide range of bulk and heterogeneous environments.

\begin{acknowledgements}
We greatly thank Louis Streacker and Dor Ben-Amotz for providing the experimental Raman spectra. This material is based upon work supported by the National Science Foundation under Grant No. CHE-1652960. T.E.M also acknowledges support from a Cottrell Scholarship from the Research Corporation for Science Advancement. This research used resources of the National Energy Research Scientific Computing Center, a DOE Office of Science User Facility supported by the Office of Science of the U.S. Department of Energy under Contract No. DE-AC02-05CH11231. We would also like to thank Stanford University and the Stanford Research Computing Center for providing computational resources and support that have contributed to these research results.
\end{acknowledgements}

\end{document}